\def\kpc{ {\rm kpc} }

\def\msun{{\rm M}_{\rm \odot}}

\def\kms{ {\rm km} {\rm s}^{-1}  }

\def\sigmaT{\sigma_{\rm T}}
\def\ds{D_{\rm s}}
\def\dd{D_{\rm d}}
\def\vT{v_{\rm T}}
\def\vcirc{v_{\rm circ}}
\def\mmin{m_{\rm min}}
\def\vtran{v_{\rm \perp}}
\def\RE{R_{\rm E}}
\def\rj{r_{\rm J}}
\def\te{\langle t_e \rangle}

\def\half{{1/2}}
\def\sinfty{\lambda_\infty}
\def\rs{r_\lambda}
\def\spose#1{\hbox to 0pt{#1\hss}}
\def\lta{\mathrel{\spose{\lower 3pt\hbox{$\sim$}}
    \raise 2.0pt\hbox{$<$}}}
\def\gta{\mathrel{\spose{\lower 3pt\hbox{$\sim$}}
    \raise 2.0pt\hbox{$>$}}}

\documentstyle[12pt,aasms4,epsf,rotate]{article}

\begin{document}

\title{Brown Dwarfs, White Knights and Demons}
\author{Geza Gyuk}
\affil{S.I.S.S.A., via Beirut 2-4, 34014 Trieste, Italy}
\and
\author{N. Wyn Evans}
\affil{Theoretical Physics, Department of Physics, 1 Keble Rd, Oxford,
OX1 3NP, UK}
\and
\author{Evalyn I. Gates}
\affil{Departments of Astronomy \& Astrophysics, The
University of Chicago, Chicago, IL 60637}
\affil{Adler Planetarium, 1300 Lake Shore Drive, Chicago, IL 60605}

\begin{abstract} 
This paper investigates the hypothesis that the lensing objects
towards the Large Magellanic Cloud (LMC) are brown dwarfs by analysing
the effects of velocity anisotropy on the inferred microlensing
masses.  To reduce the masses, the transverse velocity of the lenses
with respect to the microlensing tube must be minimised. In the outer
halo, radial anisotropy is best for doing this; closer to the solar
circle, azimuthal anisotropy is best. By using a constraint on the
total kinetic energy of the tracer population from the Jeans
equations, the microlensing mass is minimised over orientations of the
velocity dispersion tensor. This minimum mass is $\approx 0.1\,\msun$,
which lies above the hydrogen burning limit. This demonstrates
explicitly that populations of brown dwarfs with smoothly decreasing
densities and dynamically mixed velocity distributions cannot be
responsible for the microlensing events. Brown dwarfs are no white
knights!  There is one caveat.  If there are demons sitting on the
microlensing tube, they can drop brown dwarfs so as to reproduce the
microlensing data-set exactly. Such a distribution is not smooth and
does not give well-mixed velocities in phase space. It is a
permissible solution only if the outer halo is dynamically young and
lumpy. In such a case, theorists cannot rule out brown dwarfs. Only
exorcists can!

\end{abstract}


\section{INTRODUCTION}

The MACHO collaboration has interpreted its observations of
microlensing events towards the Large Magellanic Cloud (LMC) as
evidence that about one third of the halo of our own halo exists in
the form of objects of around $0.5$ solar
mass~(\cite{al97a}). Unfortunately, there are seemingly insuperable
objections to all the obvious candidates for the lensing
population. Normal stars would be visible~(\cite{al97a}), white dwarfs
are ruled out by current population II abundance
ratios~(\cite{fms,gm}), while the Hubble Deep Field gives stringent
restrictions on the contribution of red dwarfs~(\cite{gf}). The
microlensing events would be easier to understand if the
characteristic mass of the lensing objects was below the
hydrogen-burning limit ($\approx 0.08 \msun$). Of course, a lensing
population of brown dwarfs would be much too dark to be visible and
there is no conflict either with the metallicity data or the Hubble
Deep Field star counts. So, it is natural to ask the questions: Can
the deflectors be brown dwarfs? Is it possible that the masses of the
microlenses have hitherto been overestimated? The aim of this Letter
is to answer these questions.

Uncertainties in estimates of the lens candidates arise from two
fundamental sources: low number statistics and modelling
error. Although the number of microlensing events observed towards the
LMC is still low, a determination of the average mass for a {\em
given} halo model can be obtained with perhaps 50\% accuracy (see
eg. Mao \& Paczy\'nski 1996; Alcock et al. 1997). This number is
expected to improve substantially over the course of the next few
years as new events are detected. A much more important source of
error comes from our ignorance of the structure of the outer Milky Way
halo.  The halo models used by Alcock et al. (1997) are either
isotropic, such as the cored isothermal sphere~(\cite{kim}), or they
are very nearly so, such as the power-law models~(\cite{wyn}).  Alcock
et al. (1997; see especially Figures 17 and 24) plot likelihood
contours in the plane of the lens mass and baryon fraction of the
halo. The striking elongation of the contours along the baryon
fraction axis suggests that there is comparatively little uncertainty
in the mass estimates of the microlenses for a given model.  One worry
is that this propitious state of affairs is a consequence of using
halo models that all pretty much look the same!  In a percipient
investigation, Markovi\'c \& Sommer-Larsen (1997) looked at a wider
range of halos, including some with anisotropic velocity
distributions. They found that $\sim 100$ events (an order of
magnitude more than presently available) are needed to estimate the
average mass. This large error bar includes both the modelling and the
statistical uncertainty. The claims of Markovi\'c \& Sommer-Larsen
(1997) may be somewhat overstated because of the uniform priors used
in their Monte Carlo simulations. However, Mao \& Paczy\'nski (1996)
have also emphasised the difficulty of drawing firm conclusions about
the mass distribution of the lenses from the limited sample
available. These two papers consider both the statistical and the
modelling uncertainties together. The focus of our paper is on the
modelling uncertainty alone.  Our aim is to demonstrate unambiguously
that the modelling uncertainties cannot be responsible for the high
average mass estimates of Alcock et al. (1997).

Low mass lenses such as brown dwarfs are already ruled only for halo
models with negligible rotation and isotropic velocity dispersions
(e.g., Chabrier, Segretailn \& M\'era 1996). To rule out the
hypothesis that the lenses are brown dwarfs requires a thorough
investigation of halo models with very different kinematics -- in
particular with different streaming velocities and different random
motions.  Gyuk \& Gates (1998) have already shown that rotating halos
are unable to reduce the microlensing mass estimates below about
$0.25\,\msun$ (unless all the lensing takes place very close to the
Sun). This Letter will examine the effects of anisotropy and show that
the associated {\em modeling uncertainties} cannot cause the high lens
mass estimates.

\section{VELOCITY ANISOTROPY AND THE MINIMUM MASS OF THE 
MICROLENSING OBJECTS}

Let us start with a thought experiment. Suppose a stationary observer
views a stationary source through a population of lenses
with density $\rho$. The timescale of any lensing event is related to
the Einstein radius $\RE$ and the transverse velocity $\vT$ by
\begin{equation}
t_0 = {\RE \over \vT} = {1\over \vT}\sqrt{4GM\dd (\ds - \dd) \over
c^2 \ds},
\end{equation}
where $M$ is the mass of the lens and $\dd$ and $\ds$ are the distances to
deflector and source.  Suppose now that the distribution of transverse
velocities of the lenses is Gaussian with a dispersion $\sigmaT$. The
microlensing optical depth $\tau$ is well-known to be independent of the
masses of the lenses~(\cite{pg}).  The rate of microlensing $\Gamma$ is
(e.g.,~\cite{kim})
\begin{equation}
\Gamma = (2\pi)^\half {\sigmaT\over M^\half}
         \int_0^{\ds} d\dd \rho(\dd) \sqrt{4G\dd(\ds-\dd)\over c^2 \ds},
\end{equation}
and the timescale histogram is
\begin{equation}
{d \Gamma \over d t_0} = {8A^2 \sigmaT^2\over M}\int_0^{\ds} d\dd
\rho(\dd) \dd^2 (\ds - \dd)^2
\exp\Bigl[ - A\dd(\ds - \dd) \Bigr],
\end{equation}
with $A = 2 G M / (\ds c^2 t_0 \sigmaT^2)$.  This demonstrates explicitly
that all the microlensing quantities ($\tau, \Gamma, {d \Gamma \over d
t_0}$) depend only on the ratio $M/\sigmaT^2$. Given the microlensing
data-set alone, it is not possible to constrain the mass of the deflectors
at all! Any mass estimate is solely a consequence of assumptions regarding
the transverse velocity dispersions. The same data will be consistent with
smaller inferred mass if the transverse motions are reduced.  This
degeneracy between mass and velocity can be partially lifted if parallax
effects~(\cite{sjur,kim,andy}), or finite source size
effects~(\cite{nemiroff}) can be detected in the lightcurve.  

Of course, the analysis of the microlensing events towards the LMC is
more complex than this thought experiment. Both the Sun and the LMC
are moving and therefore the expectation value of the transverse
velocity of the lens with respect to the microlensing tube cannot be
made arbitrarily small just by changing the velocity anisotropy of the
lenses. Figure 1 shows a planform of the Sun and the LMC projected on
the Galactic equatorial plane. The line of sight from the Sun to the
LMC is shown as a dashed line. In the outer parts of the halo, this
line of sight is aligned very nearly with the radial direction of the
spherical polar coordinate systems.  Radial anisotropy of the velocity
dispersion tensor is the best option for reducing the mass estimates.
Nearer the solar circle, the offset of the Sun from the Galactic
Center becomes important. The line of sight is aligned more nearly
with the azimuthal direction. This means that radial anisotropy is now
dangerous. The best recipe for the minimum microlensing mass is to
allow the velocity dispersion tensor to be azimuthally distended near
the Sun and to become radially distended in the outer halo.

As a simple model, let us assume that the density of the lensing
population is smooth and falls off like a power of the distance ($\rho
\propto r^{-\gamma}$). We take the overall potential to be a power-law model,
so that the circular velocity, $\vcirc$, falls like $r^{-\beta/2}$.  Rich
families of solutions to the Jeans equations for power-law density
distributions in power-law potentials are known~(\cite{ehz}).  These are
all aligned in the spherical polar coordinates, but vary in the anisotropy
of the principal components of the velocity dispersion tensor
$\sigma_i$. The detailed Jeans solutions all satisfy the constraint (see
eq. (3.8) of
\cite{ehz})
\begin{equation}
\sum_{i=1}^3 \sigma^2_i \gta {\vcirc^2} \times 
{\rm min} \Bigr( 1,  {1 \over \beta + \gamma -2 } \Bigl).
\label{eqone}
\end{equation}
From the standpoint of minimising the microlensing mass estimates, the
best of all possible worlds is to replace the inequality in the above
expression with an equality. This means that the total kinetic energy
required to support the lensing population against gravity is
underestimated. The inferred microlensing mass will always be lower than
the true mass.  We allow the ratio of the principal components of the
velocity dispersions to vary subject only to the condition that the
sum of the components does not violate the inequality (4). Thus, the
Jeans equations are not satisfied spot-wise, but only in a gross
sense.  The total kinetic energy cannot be reduced further without
violating the rules of gravitational physics.  If all the deflectors
are $1 \msun$ objects, the rate is (see e.g., \cite{kim,gg})
\begin{equation}
\Gamma = 2\int_0^{\ds} d\dd \RE (\dd) \rho(\dd) \langle |\vtran|
\rangle.
\end{equation}
Here $\langle | \vtran | \rangle$ is the average value of the
transverse velocity of the lens with respect to the microlensing tube.
The best estimator of the average event duration $\te$ is 61 days (see
Appendix A of~\cite{gg}). This uses the events and the efficiencies
given in Alcock et al. (1997a). So the microlensing mass estimate
$\mmin$ is
\begin{equation}
\mmin \gta \Biggl[ \te {\Gamma \over \tau} \Biggr]^2.
\end{equation}
Our algorithm for finding the minimum microlensing mass estimate is as
follows. Choose the model parameters $\beta$ and $\gamma$ and an
alignment of the velocity dispersion tensor, and then apportion the
total kinetic energy into the three principal components subject only
to the constraint (4). The velocity anisotropy may be parametrised by
\begin{equation}
\lambda = {{\sigma_2^2 + \sigma_3^2} \over {2 \sigma_1^2}}, \qquad\qquad \mu = {\sigma_3^2\over
\sigma_2^2}.
\end{equation}
Let us insist that the velocity ellipsoid cannot be anisotropic by
more than a $4:1$ ratio; that is to say, $\lambda$ and $\mu$ must lie
within the range $1/16$ to $16$.  For comparison, Freeman (1987)
reports that the Population II stars in the spheroid have velocity
dispersions at the solar position oriented on the cylindrical polar
coordinate system such that $(\sigma_R, \sigma_\phi, \sigma_z) \approx
(140, 100, 75)\,\kms$. The microlensing mass is to be minimised as
both the anisotropy parameters $\lambda$ and $\mu$ and the alignment
of the velocity ellipsoid are varied.

After a little thought, it is obvious what the alignment is for the
minimum mass estimate -- the best of all possible worlds is when the
velocity ellipsoid is aligned along the microlensing tube itself, with
as much kinetic energy as possible put into motions along the tube and
as little as possible put into transverse motions.  Figure 2 shows the
inferred mass as a function of $\gamma$ where the mass has been
minimized by allowing $\lambda$ and $\mu$ to float over their
respective ranges. Different curves correspond to the range $-0.25 \le
\beta \le 0.25$. In such a situation, $\lambda$ always prefers to be as
low as possible, while $\mu$ is only weakly constrained. The minimum
mass estimate ranges from $0.1\, \msun$ when $\gamma =2.0$ to $0.25\,
\msun$ when $\gamma =4.0$. The inferred mass is always larger than the
hydrogen-burning limit. This leads us to the main result of the
Letter.  {\it If the density of the microlensing population is smooth
and monotonic decreasing (that is, reasonably well-approximated by a
power-law), then the microlenses cannot be brown dwarfs, irrespective
of the details of their kinematics}. The strength of this statement is
that it is based on the Jeans equations and therefore robust.

\section{A MODEST ESTIMATE}

Let us emphasise that this algorithm for obtaining the minimum mass gives
a value that is {\em very much a lower limit}. It uses a number of gratuitous
approximations, all of which act to reduce the mass estimate. For example,
the Jeans solutions of reasonable tracer populations may possess a
kinetic energy greater than the minimum prescribed by eq. (4).  Again,
almost certainly, the alignment along the microlensing tube that yields
the minimum mass cannot be built -- that is, there is no set of stellar
orbits that can be superposed to yield a true dynamical model
corresponding to the Jeans solution.  Making the model more realistic will
necessarily require more massive lenses. In this section we provide an
estimate of the more modest reduction in the microlensing masses expected
from velocity anisotropy for one particular reasonably realistic model
of the halo.

To do this, let us build Jeans solutions of tracer populations with
the density of the Jaffe (1983) sphere
\begin{equation}
\rho = {M\over 4 \pi \rj^3}{\rj^4\over r^2( r+ \rj)^2}
\end{equation}
in a spherical isothermal halo potential. Here, $\rj$ is a
scale-length that describes when the density turns over. A typical
estimate of its value might be $\rj \sim 50 \kpc$
(\cite{chris,markw}). Let the anisotropy be defined as
\begin{equation}
{\sigma_\theta^2\over \sigma_r^2} = {\sigma_\phi^2 \over \sigma_r^2}
= \sinfty + {\rs \over r} .
\end{equation}
This simple ansatz allows the kinematics to change from azimuthal
anisotropy to radial anisotropy or vice versa. Here, $\sinfty$ is
the value of the anisotropy at infinity, whereas $\rs$ is a
scale-length on which the anisotropy changes. The solution of the
spherical Jeans equation is readily found by the method of integrating
factors as (see Binney \& Tremaine 1987)
\begin{equation}
\sigma_r^2 = \vcirc^2 \exp(-2\rs /r) r^{2\sinfty} (r + \rj)^2
             \int_r^\infty {dr \exp(2\rs /r) \over r^{2\sinfty +1}
             (r+ \rj)^2}.
\end{equation}
The azimuthal dispersions now follow from (9).  The isotropic model
has $\sinfty =1$ and $\rs =0$ and a mass estimate (given by eq. 6) of
$0.348\,\msun$. This can be reduced by anisotropy. The microlensing
mass estimate in the plane of the asymptotic anisotropy $\sinfty$ and
the anisotropy scale $\rs$ is shown in Figure 3. For this set of Jeans
solutions -- in which the anisotropy can change significantly but not
dramatically -- there is no hope of using anisotropy by itself to
reduce the microlensing mass estimate below $\approx 0.3\,\msun$.

\section{DISCUSSION AND CONCLUSIONS}

If the density of the microlenses is smooth and decreasing, then they
cannot be brown dwarfs. This holds irrespective of the details of
their kinematics. This general result follows because the Jeans
equations (or, equivalently, the virial theorem) imply the existence
of an irreducible minimum kinetic energy to support the lensing
population against gravity. Even in the optimum alignment of the
velocity dispersion tensor of the lenses, this must yield sufficient
transverse motion so that the minimum mass is $\approx 0.1\,\msun$ for
halo models with flat rotation curves. This is above the hydrogen
burning limit. 

There is a way to save brown dwarfs.  Let us imagine a collection of
demons sitting on the microlensing tube. One of the demons at a
heliocentric distance of 20 kpc launches a brown dwarf of mass
$0.06\,\msun$ with a velocity of $106\, \kms$ across our line of sight
... and this causes event \# 4 with a blended timescale of 39.5 days. A
second demon sitting on the microlensing tube at 30 kpc lobs a brown dwarf
with a velocity of just $75 \,\kms$ ... and this gives event \# 5 with a
blended timescale of 55.5 days, and so on. Of course, demons can exactly
reproduce the dataset reported by Alcock et al. (1997) by dropping brown
dwarfs from the microlensing tube.  The density of brown dwarfs so
produced is neither spherical nor axisymmetric nor in a steady-state. The
velocity distribution is not dynamically well-mixed and the time averages
theorem (Binney \& Tremaine 1987, p. 171), which is the fundamental result
underpinning steady-state stellar dynamics, does not hold. If it did, we
could infer the existence of further brown dwarfs at different phases of
the same orbits and show that they produce microlensing events that are
not seen. Such a model is possible if the halo is very blobby
(e.g.~\cite{donald,donaldruth}). Then, in every direction that one looks
(including $\ell = 280^\circ, b = -33^\circ$), there may be garbage heaps
of brown dwarfs whose density and velocity distributions are lumpy. This
possibility cannot be ruled out from the microlensing data-set
alone.

\acknowledgments

Most of this work was done during a visit to SISSA, Trieste. NWE
wishes to thank the Astrophysics Sector in general, and Dennis Sciama
and John Miller in particular, for their kindness and their
hospitality. NWE thanks James Binney for numerous insightful remarks
on this subject.

\eject

\begin{figure}
\begin{center}
              \epsfxsize 0.5\hsize
               \leavevmode\epsffile{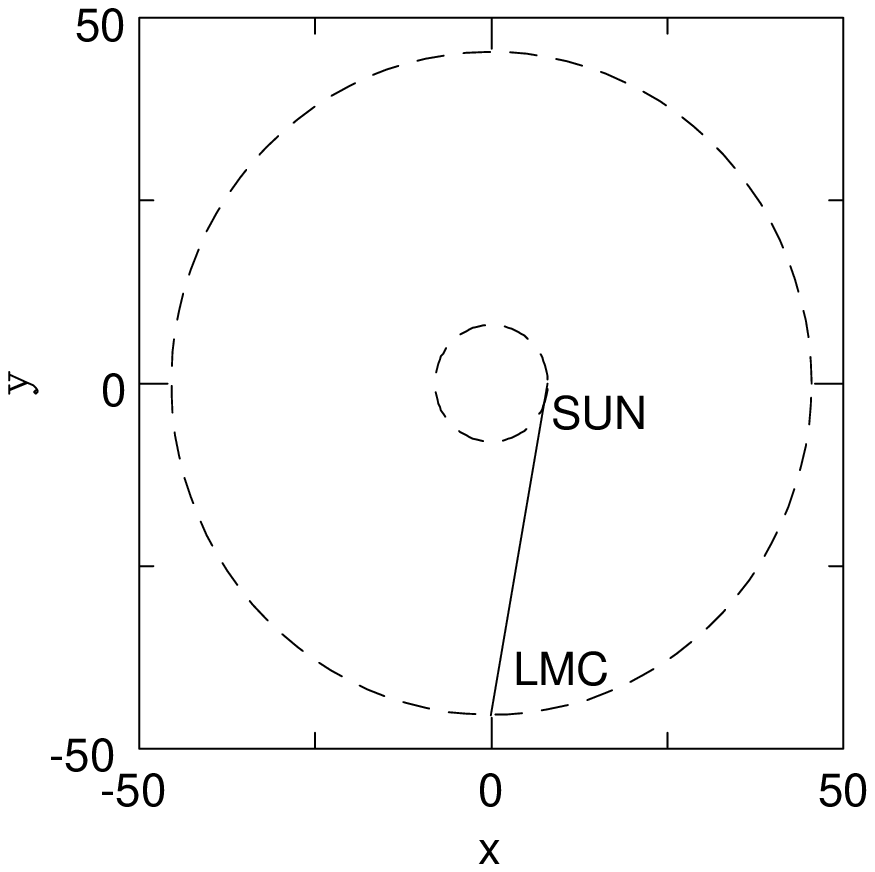}
\caption{The positions of the Sun and the Large Magellanic Cloud (LMC)
projected onto the Galactic Plane. The line of sight from the Sun
to the LMC is marked. This is almost radially aligned in the outer
halo, but azimuthally aligned near the Solar circle.}
\end{center}
\end{figure}

\eject

\begin{figure}
\begin{center}
              \epsfxsize 0.5\hsize
               \leavevmode\rotate[r]{\epsffile{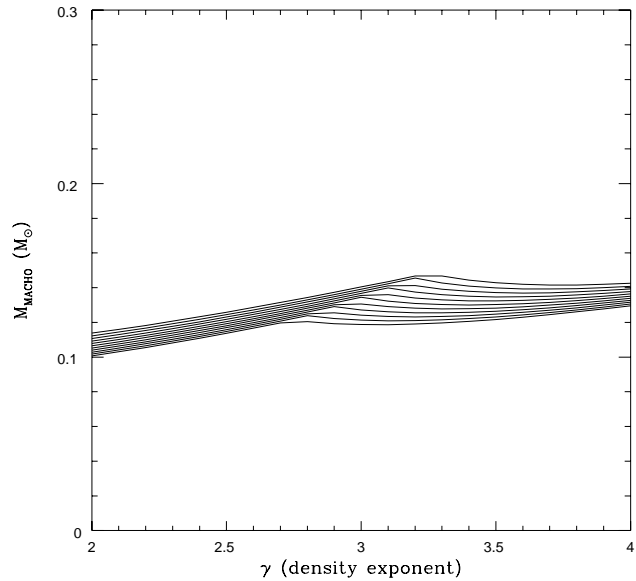}}
\caption{The minimum mass estimate as a function of the steepness of the
lens density distribution, $\gamma$. As described in the text both
anisotropies have been allowed to float. The break in the curve at
$\gamma=3$ is due to the prescription for finding the minimum kinetic
energy. For all $\gamma$ the microlensing mass is $\gta 0.1\,\msun$,
which is too massive for cold, degenerate brown dwarfs. While
recognising the statistical uncertainties are still great, the virtue
of this figure is that it demonstrates that the modelling
uncertainties associated with anisotropy cannot be responsible for the
high lens mass estimates. }
\end{center}
\end{figure}

\begin{figure}
\begin{center}
              \epsfxsize 0.5\hsize
               \leavevmode\epsffile{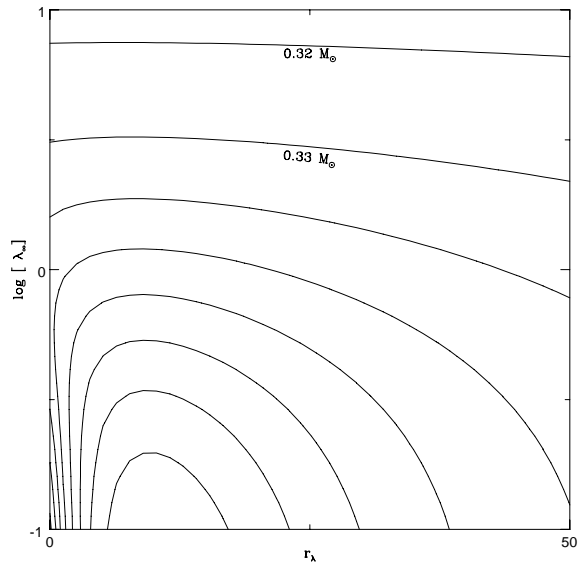}
\caption{Contour plots of the minimum microlensing mass for the
Jaffe model in the plane of the model parameters $(r_\lambda, \log
\lambda_\infty$). The contours are spaced at intervals of
$0.01\,\msun$. Anisotropy does change the mass estimate --
but never below $\approx 0.3 \,\msun$. For this particular Jaffe model
with $\rj \sim 50\,\kpc$, radial anisotropy is best for reducing
the mass estimate.}
\end{center}
\end{figure}

\end{document}